\documentclass[aps,superscriptaddress,twocolumn,twoside,floatfix,prb]{revtex4-2}
\usepackage{times}
\usepackage{epsfig}
\usepackage{amsfonts}
\usepackage{amsmath}
\usepackage{amssymb,amsthm}
\usepackage{xcolor,colortbl}
\usepackage{multirow}
\usepackage{braket}
\usepackage{latexsym}
\usepackage{amsfonts}
\usepackage{mathrsfs}
\usepackage{natbib}
\usepackage{verbatim}
\usepackage{gensymb}
\usepackage{caption}
\usepackage{ragged2e}
\DeclareCaptionJustification{justified}{\justifying}
\usepackage{blkarray}
\usepackage{graphicx}
\usepackage{enumerate}
\usepackage[shortlabels]{enumitem}
\usepackage{ulem}
\usepackage{chemformula}
\usepackage{soul}

\usepackage[colorlinks=true,linkcolor=red,citecolor=blue,urlcolor=blue]{hyperref}
\allowdisplaybreaks

\newcommand{\supp}{See supplementary material at [url] for additional details on sample, preparation, sample characterization, measurement details, fitting methodology and other calculations}
\begin{document}

\title{Effects of Reduced Interlayer Interactions on the K-point Excitons of \ch{MoS2} Nanoscrolls}

\author{Sagnik \surname{Chatterjee}}
\email{chatterjee.sagnik@students.iiserpune.ac.in}
\affiliation{Department of Physics, Indian Institute of Science Education and Research (IISER), Pune 411008, India}
\author{Tamaghna \surname{Chowdhury}}
\email{tamaghna.chowdhury@students.iiserpune.ac.in}
\affiliation{Department of Physics, Indian Institute of Science Education and Research (IISER), Pune 411008, India}
\affiliation{Department of Physics and Astronomy, University of Manchester, United Kingdom M13 9PL}
\author{Pablo \surname{D\'iaz-N\'u\~nez}}
\affiliation{Department of Physics and Astronomy, University of Manchester, United Kingdom M13 9PL}
\affiliation{National Graphene Institute, University of Manchester, United Kingdom, M13 9PL.}
\author{Nicholas \surname{Kay}}
\affiliation{Department of Physics and Astronomy, University of Manchester, United Kingdom M13 9PL}
\affiliation{National Graphene Institute, University of Manchester, United Kingdom, M13 9PL.}
\author{Manisha \surname{Rajput}}
\affiliation{Department of Physics, Indian Institute of Science Education and Research (IISER), Pune 411008, India}
\author{Sooyeon \surname{Hwang}}
\affiliation{Center for Functional Nanomaterials, Brookhaven National Laboratory, Upton, NY 11973, USA }
\author{Ivan \surname{Timokhin}}
\affiliation{Department of Physics and Astronomy, University of Manchester, United Kingdom M13 9PL}
\affiliation{National Graphene Institute, University of Manchester, United Kingdom, M13 9PL.}
\author{Artem \surname{Mishchenko}}
\email{artem.mishchenko@gmail.com}
\affiliation{Department of Physics and Astronomy, University of Manchester, United Kingdom M13 9PL}
\affiliation{National Graphene Institute, University of Manchester, United Kingdom, M13 9PL.}
\author{Atikur \surname{Rahman}}
\email{atikur@iiserpune.ac.in}
\affiliation{Department of Physics, Indian Institute of Science Education and Research (IISER), Pune 411008, India}

\date{\today}
\begin{abstract}
Transition metal dichalcogenide (TMD) nanoscrolls (NS) exhibit significant photoluminescence (PL) signals despite their multilayer structure, which cannot be explained by the strained multilayer description of NS. Here, we investigate the interlayer interactions in NS to address this discrepancy. The reduction of interlayer interactions in NS is attributed to two factors: (1) the symmetry-broken mixed stacking order between neighbouring layers due to misalignment, and (2) the high inhomogeneity in the strain landscape resulting from the unique Archimedean spiral-like geometry with positive eccentricity. These were confirmed through transmission electron microscopy, field emission scanning electron microscopy and atomic force microscopy. To probe the effect of reduction of interlayer interactions in multilayered \ch{MoS2} nanoscrolls, low-temperature PL spectroscopy was employed investigating the behaviour of K-point excitons. The effects of reduced interlayer interactions on exciton-phonon coupling (EXPC), exciton energy, and exciton oscillator strength are discussed, providing insights into the unique properties of TMD nanoscrolls. 
\end{abstract}

\maketitle

\def\ans{X_{NS}^A}
\def\bns{X_{NS}^B}
\def\aml{X_{ML}^A}
\def\bml{X_{ML}^B}
\def\ians{I_{X_{NS}^A}}
\def\ibns{I_{X_{NS}^B}}
\def\iaml{I_{X_{ML}^A}}
\def\ibml{I_{X_{ML}^B}}
\def\a{X^A}
\def\b{X^B}
\def\ia{I_{X^A}}
\def\ib{I_{X^B}}
\def\ms{\mathrm{MoS}_2}

2D-TMDs are known for their diverse and exceptional optoelectronic properties, which can be tuned using various techniques such as applying strain, modifying their dielectric environment, or fabricating stacked heterostructures with twist angles \cite{makprl,splendiani2010emerging,chowdhury2021modulation,PhysRevB.98.115308}. Among these strategies,  scrolling up TMD monolayers (ML) into quasi-one-dimensional systems, known as TMD nanoscrolls (NS), has emerged as a compelling method to create structures with distinctive features, including large effective surface area, unique self-encapsulated geometry, and multilayer tubular structure \cite{cui2018rolling}. 

The recent breakthrough in the facile synthesis of these TMD NS has sparked a cascade of efforts within the scientific community to explore and exploit the potential applications of these materials across diverse fields \cite{cui2018rolling,wang2022strong,qian2020chirality,doi:10.1021/acsnano.3c05681,liu2017two, qiao2024one, Zhu2024}.
The distinctive structure of NS, coupled with their emerging PL, makes them a compelling platform for investigating interlayer interactions (ILI) and their impact on the excitonic properties of NS. The cylindrical curved geometry and layer stacking greatly influence the ILI of TMDs \cite{koskinen2014density,huang2014probing,liu2014evolution,grzeszczyk2021optical}.  ILI play a pivotal role in shaping the optoelectronic properties of multilayer TMDs by influencing the electronic bandstructure, for instance, introducing band-splitting, or direct to indirect bandgap transition\cite{splendiani2010emerging,makprl,PhysRevLett.111.106801}. The valence band near the K-point splits due to spin-orbit coupling (SOC) and ILI in TMDs giving rise to two K-point excitons: A-exciton ($\a$) and B-exciton ($\b$) \cite{zhang2015valence}. Thus, PL spectroscopic investigation can reveal the nature of SOC and ILI in multilayered NS. \par

\begin{figure}[b]
    \centering
    \includegraphics[width = 0.92\linewidth]{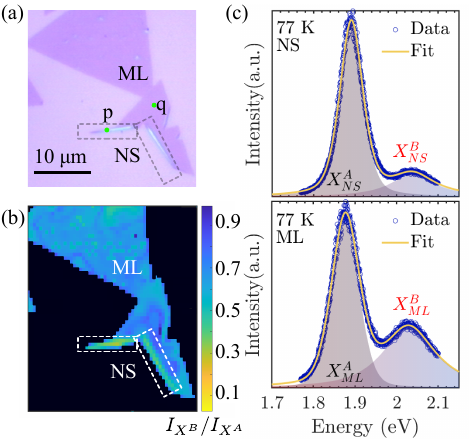}
   \caption{(a) Optical image of partially scrolled \ch{MoS2} ML. Green spots denoted by p and q are excitation spots for PL measurements. The diameter of the spots p and q matches exactly with the laser spot size. (b)Mapping of $\ib/\ia$ at room temperature for a partially scrolled \ch{MoS2} ML. Scrolled regions at the edges are highlighted with a box in (a) and (b). (c) Low-temperature PL spectra of \ch{MoS2} NS and ML. $\a$ and $\b$ peaks are fitted with two Lorentzian functions (highlighted area showing the individual peaks)}
    \label{fig:1}
\end{figure}
In this study, we have investigated the ILI in multilayered \ch{MoS2} NS by examining the behaviour of K-point excitons through PL spectroscopy. Additionally, we employed transmission electron microscopy (TEM) to explore potential stacking orders within the \ch{MoS2} NS layers. Our TEM results reveal the presence of mutually twisted layers and broken inversion symmetry within the NS structure. These findings prompt a deeper discussion of the origins of our spectroscopic observations. Specifically, we report lowered ILI and the presence of layer decoupling in the NS system, which we correlate with its cylindrical curved geometry and stacking. Furthermore, we discuss the influence of layer decoupling on several key properties of K-point excitons in NS, including EXPC, exciton energy, and exciton oscillator strength. In numerous preceding studies, NS have demonstrated their excellence as photodetectors \cite{fang2018transforming,doi:10.1021/acsami.1c24291,deng2019high,doi:10.1021/acs.jpclett.0c00861}; given that excitonic properties largely influence the optoelectronic response, our investigation assumes a crucial role in enhancing the comprehension and achieving better modulation in the application of NS as optoelectronic devices.

The PL measurements were performed using a continuous wave laser of wavelength 514.5 nm (2.41 eV) on the sample shown in FIG. \ref{fig:1}(a) (see supplementary section I for measurement details ~\footnote{\supp}). FIG. \ref{fig:1}(b) shows the map of the ratio of $\b$ peak intensity ($\ib$) and $\a$ peak intensity ($\ia$) for a partially scrolled NS at ambient conditions (see supplementary section II for the map of the peak position of $\a$ and $\b$). The $\ib/\ia$ value in the NS is significantly lower than in the ML-\ch{MoS2}, which is unexpected given the multilayer structure of the NS. Previous reports have shown that as the number of layers increases, $\ia$ decreases and becomes comparable to $\ib$, causing $\ib/\ia$ value to approach unity\cite{zhang2015valence,splendiani2010emerging}. The mapping was done with an objective lens of 100$\times$ magnification and a high numerical aperture (N. A.) $=$ 0.95 which gives a diffraction-limited laser spot size of $\sim$ 660 nm (see supplementary section I for spot size calculation ~\footnotemark[\value{footnote}]). To avoid flake-to-flake variation, we have compared the $\ib/\ia$ value for scrolled and un-scrolled parts of the same \ch{MoS2} ML [FIG.  \ref{fig:1}(a)].  

To better understand these observations, we measured PL at a low temperature of 77 K and pressure of $2\times10^{-5}$ mbar. The measurement was done with a 100$\times$ objective lens of N. A. = 0.80 which gives a diffraction-limited laser spot size of $\sim$ 800 nm. To collect the signal solely from the NS (ML) and to avoid the contribution of ML (NS), measurements were done at the spot ``p" (``q") as shown in FIG. \ref{fig:1}(a). FIG.  \ref{fig:1}(c) shows the normalized PL spectra of NS and ML at 77 K. The lineshape is fitted with two Lorentzian functions corresponding to the spin-orbit split excitonic peaks, $\a$ and $\b$, for both \ch{MoS2} NS and ML. The peak positions for $\ans$, $\bns$, $\aml$, and $\bml$ are 1.89 eV, 2.04 eV, 1.87 eV, and 2.03 eV, respectively. Compared to the ML, the $\a$ and $\b$ peaks in the NS exhibit a blueshift of approximately 20 meV and 10 meV, respectively. Furthermore, significant suppression $\bns$  is clearly visible in the PL spectrum.

 We also performed temperature-dependent PL measurements on NS and ML at the earlier mentioned locations (FIG. \ref{fig:2}(a, b)). The energy of $\a$ and $\b$ excitons ranges between $\sim$ 1.81$-$1.87 eV and 1.93$-$2.05 eV for various temperatures. The temperature-dependent ratio $\ib/\ia$ for ML consistently remains higher than that of the NS [FIG.  \ref{fig:2}(c)]. The energy difference $E_{\b}-E_{\a}$ is a measure of the valence band maxima splitting ($\Delta_{vb}$) due to SOC and ILI. However, for the ML, only SOC contributes to the splitting. For multilayer ($>$ 2L) \ch{MoS2}, the temperature-dependent $\Delta_{vb}$ can be well approximated by the following expression\cite{vina1984temperature,zhang2015valence}: 

\begin{figure}[t]
    \centering
    \includegraphics[width = 0.85\linewidth]{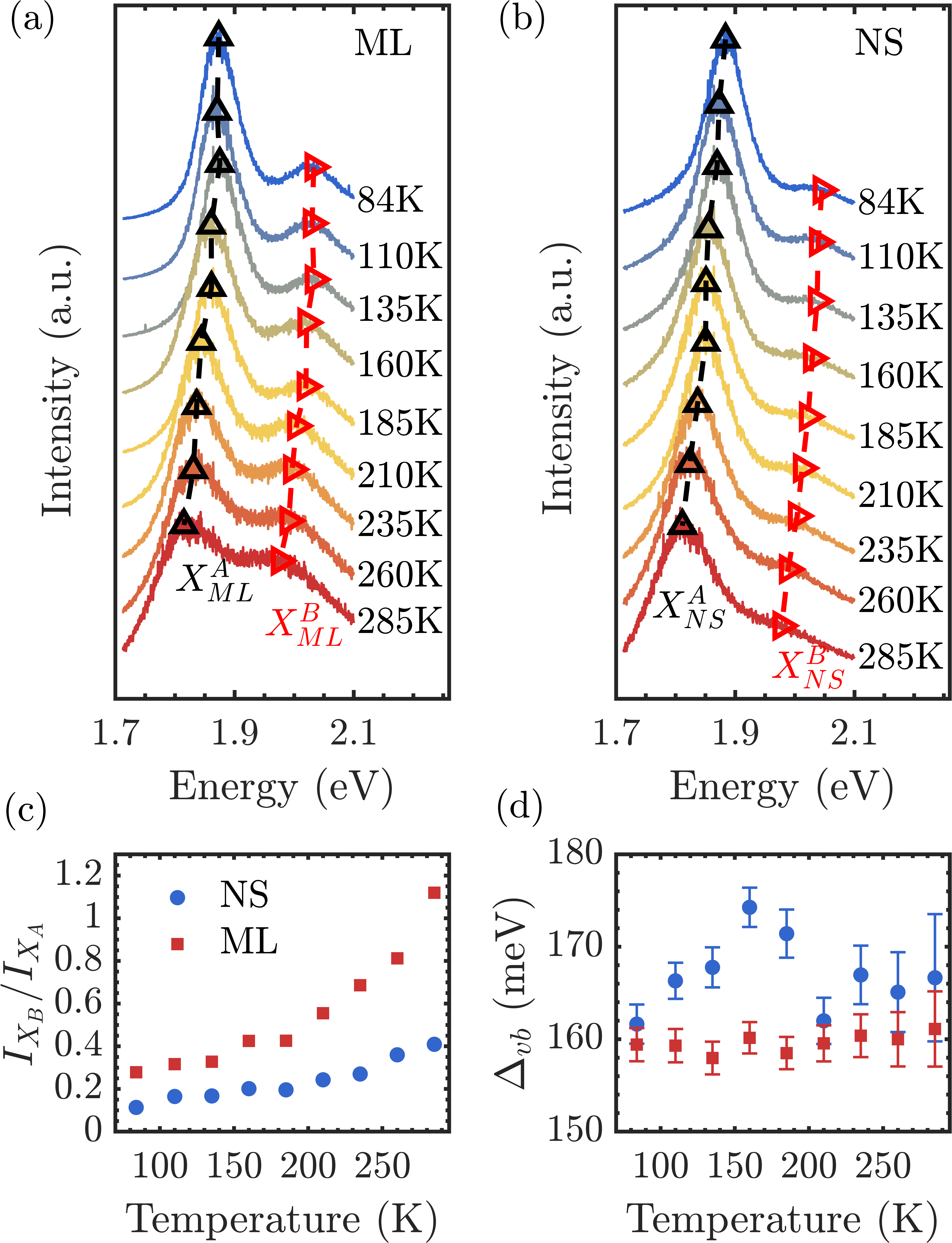}
    \caption{Temperature Dependent PL Spectra for (a) ML (b) NS. Triangular markers show the peak positions for $\a$ (black) and $\b$ (red). (c) Temperature dependence of $\ib/\ia$ for ML and NS. (d) Temperature dependence of $\Delta_{vb}$ for ML and NS. Blue and red circles correspond to NS and ML respectively.}
    \label{fig:2}
\end{figure}

\begin{equation}
	\Delta_{vb}(T) = \Delta_{vb}(0) - \frac{\alpha}{e^{{\frac{\Theta}{K_BT}}}-1}
    \label{eq:1}
\end{equation}

Here, $\Delta_{vb}(0)$ is the value of $\Delta_{vb}(T)$ at T = 0 K; $\alpha$ and $\Theta$ are the fitting parameters.$\alpha$ is related to the ILI strength, and $\Theta$ can be correlated to the energy of the out-of-plane transverse optical (TO) phonon \cite{zhang2015valence}. For ML, the value of $\alpha = 0$, which means $\Delta_{vb}(0)$ does not exhibit any temperature dependence. For thicker layers ($\geq$ 2), the value of $\alpha$ increases due to increasing ILI \cite{PhysRevLett.111.106801,zhang2015valence}. Consequently, $\Delta_{vb}$ shows a strong temperature dependence for thicker layers\cite{zhang2015valence}. Considering NS, the number of layers in each NS must be greater than 2. Therefore, $\Delta_{vb}^{NS}$ should show a strong temperature dependence. However, from Figure 2d, we can observe that although the value of  $\Delta_{vb}^{NS}$ is higher than that of the ML, it does not show any significant temperature dependence [FIG.  \ref{fig:2}(d)]. This observation indicates the presence of only weak ILI in the NS.

\begin{figure}[h]
    \centering
    \includegraphics[width = \linewidth]{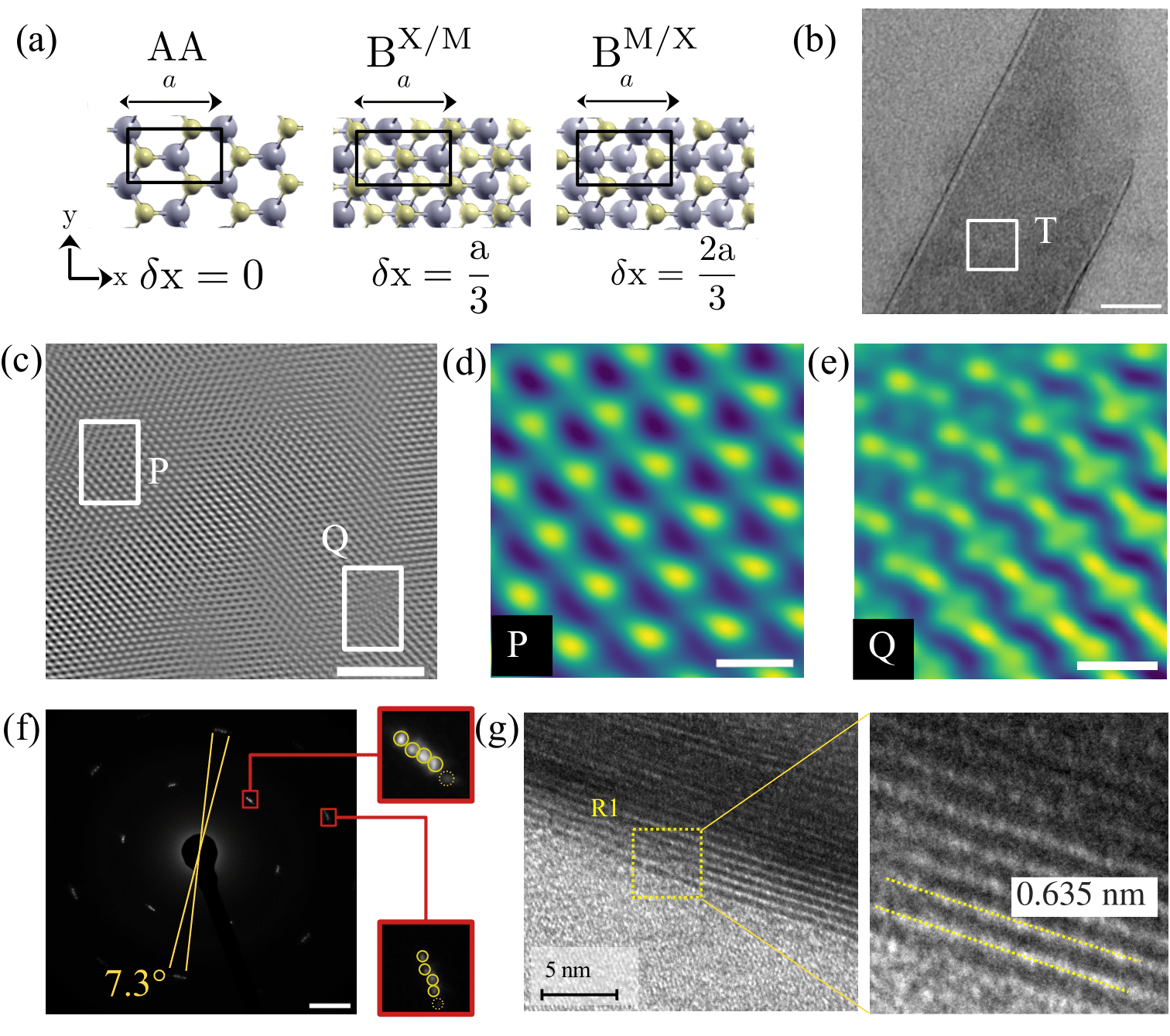}
   \caption{(a) AA, B$^{\rm{X/M}}$ and B$^{\rm{M/X}}$ stacking. The yellow and blue spheres indicate S and Mo atoms, respectively. (b) HRTEM image of the tip of an NS. Gaussian blur is applied (scale bar: 20 nm). (c) HRTEM image of the top of the NS. An inverse fast Fourier transform (IFFT) filter is applied (scale bar: 2 nm). (d) and (e) are zoomed-in images of regions P and Q marked in (b) respectively. (c) showing AA stacking and (d) showing mixed stacking (scale bar: 0.2 nm). (f) SAED taken in the region T marked in (b) showing mutually twisted layers of \ch{MoS2} with a maximum twist angle of 7.3$^{\circ}$ (scale bar: 2 nm$^{-1}$). (g) HAADF-STEM image of an edge of an NS showing the interlayer spacing of 0.635 nm. IFFT analysis is done on the selected region R1 to find out the periodicity along the perpendicular direction to the NS layers.}
    \label{fig:3}
\end{figure}

In 2H-\ch{MoS2}, there are five non-equivalent high-symmetry stacking configurations \cite{naik2019kolmogorov} (see supplementary Section III)~\footnotemark[\value{footnote}], which can be obtained by sliding the top layer with respect to the bottom layer. Starting with the AA stacking ($\delta x = \delta y = 0$), sliding the top layer by $\delta x = a/3$  leads to the B$^{\rm{X/M}}$ stacking  and sliding it by $\delta x = 2a/3$  leads to the B$^{\rm{M/X}}$ [FIG. \ref{fig:3}(a)]. 
If two parallel layers of TMDs are twisted within an angle between 0$^{\circ}$-60$^{\circ}$, intermediate stacking arrangements can be obtained that are not in a high-symmetry configuration \cite{liu2014evolution}. The ILIs decrease with increasing twist angle, with a minimum around 30$^{\circ}$\cite{huang2014probing,liu2014evolution,grzeszczyk2021optical,10.1063/5.0177357}. FIG. \ref{fig:3}(b-e) shows high-resolution transmission electron microscopy (HRTEM) images of the NS, where various mixed stacking arrangements can be observed. These arrangements originate from misaligned layers due to either sliding, twisting, or both. The selected area electron diffraction (SAED) pattern [FIG. \ref{fig:3}(f)] reveals mutually twisted layers with a maximum angular spread of 7.3$^{\circ}$ between the layers. This effective twist may vary from sample to sample as the scrolling for individual layers cannot be controlled (see supplementary section IV for HRTEM performed on another NS ~\footnotemark[\value{footnote}]). The misalignment and relative twist between layers of NS leads to broken inversion symmetry and results in weaker ILI, as observed in previous studies on twisted TMDs\cite{huang2014probing,liu2014evolution,grzeszczyk2021optical}. 
The interlayer separation of the NS observed in HAADF-STEM is $\sim$ 0.635 nm [FIG. \ref{fig:3}(g)] (see supplementary section IV for HAADF-STEM measurement details and analysis~\footnotemark[\value{footnote}]). Previous studies have reported the interlayer separation of NS to be as high as 0.65 nm\cite{fang2018transforming}, which is higher than the interlayer separation for anti-parallel AB stacked \ch{MoS2} ($\sim$ 0.6$-$0.62 nm) and match well with the interlayer separation for twisted layers ($\sim$ 0.638-0.65 nm) for twist angles between $0^{\circ}-60^{\circ}$\cite{liu2014evolution,huang2014probing,wakabayashi1975lattice,coehoorn1987electronic}. Thus this increase in interlayer separation is related to the lowered ILI in slightly twisted TMD layers. Consequently, the layers can effectively be treated as isolated MLs, and their optical responses will be an aggregation of their individual contributions.

\begin{figure}[t]
    \centering
    \includegraphics[width = 0.9\linewidth]{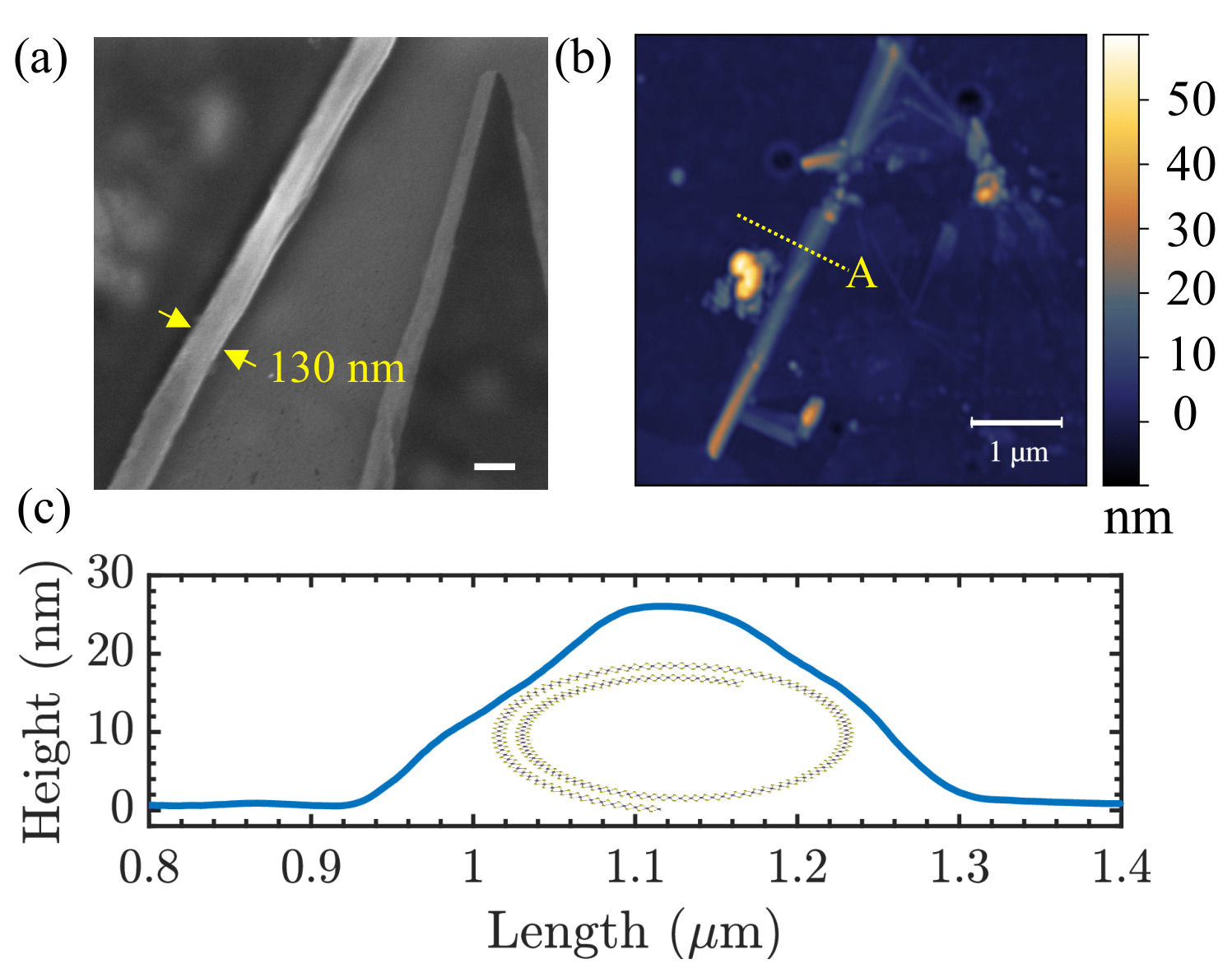}
   \caption{(a) FESEM image of partially scrolled \ch{MoS2} indicating the width of NS (scale bar: 200 nm). (b) AFM scan of partially scrolled \ch{MoS2}. (c) Height profile along line A marked in (b).}
    \label{afm}
\end{figure}
To estimate the effective ILI, the interlayer hopping parameter ($t_{\perp}$) can be calculated from $\Delta_{vb}$ as described by the $\mathbf{k \cdot p}$ model in the vicinity of K points for bilayer systems \cite{paradisanos2020controlling,slobodeniuk2019fine,grzeszczyk2021optical} as 
{\begin{equation}
    t_{\perp} = \frac{1}{2}\sqrt{\Delta_{vb}^2 - \Delta_{SO}^2}
    \label{Eq:int}
\end{equation}

Here, $\Delta_{SO}$ is the splitting of the valence band in the absence of ILI\cite{paradisanos2020controlling}, arising from intra-layer SOC, while interlayer SOC can be neglected\cite{fan2016valence,slobodeniuk2019fine}. The $\Delta_{vb}$ value for ML is a good estimation for $\Delta_{SO}$ within this framework. This framework can be extended to trilayer systems where one layer interacts with both layers above and below. In such cases,
\begin{equation}
        t_{\perp}' = \frac{1}{2\sqrt{2}}\sqrt{\Delta_{vb}^{'2} - \Delta_{SO}^2}
    \label{Eq:int3}
\end{equation}}
Therefore, the estimation of the interlayer hopping parameter using Eq. \ref{Eq:int} gives an upper limit of ILI for NS systems for the experimentally determined value of $\Delta_{vb}$ and $\Delta_{SO}$. Moreover, the interaction between the second nearest layer is an order of magnitude smaller in \ch{MoS2} for the K point\cite{fan2016valence} and can be neglected.  Earlier reports show that t$_{\perp}$ for bilayer system is $\sim 43$ meV and for even higher number of layers, it increases causing an increase in $\Delta_{vb}$\cite{shinde2018stacking, paradisanos2020controlling,gong2013magnetoelectric,fan2016valence}. For NS, the experimentally estimated value of $t_{\perp}$ of NS is 24 meV, which is significantly lower than that of a multilayer system. This reduction is in agreement with previous reports on the reduction of ILI in twisted layers of \ch{MoS2}, as described earlier.

In folded and bi-folded \ch{MoS2} monolayers and bilayers, the layer decoupling effect was previously demonstrated in ref. \cite{castellanos2014folded} which was later complemented by density functional tight binding study by Koskinen et al.\cite{koskinen2014density}. The effective layer decoupling was attributed to an inhomogeneous strain originating from the cylindrical bend. NS possess a high aspect ratio and a non-uniform curvature as confirmed by Field emission scanning electron microscopy (FESEM)(Zeiss Crossbeam 350 FIB-SEM) [FIG. \ref{afm}(a)] and atomic force microscopy (AFM) [FIG. \ref{afm}(b,c)] (see supplementary section V for details of AFM measurement~\footnotemark[\value{footnote}]) as well as from previous report in ref.\cite{qian2020chirality}. The NS surface can be effectively modelled as an Archimedean spiral with eccentricity $e>0$. Thus, the value of curvature at any point on the NS will be,
\begin{equation}
\begin{split}
     &\bigg|\frac{4\pi}{\beta^3}\bigg[h^2+2\pi^2 (r_0^m+\frac{h}{2\pi}\theta)(r_0^M+\frac{h}{2\pi}\theta) - \frac{1}{2}\pi h \sin2\theta\Delta_r\bigg]\bigg|,\\
&\text{with} \ \beta = \big[h\cos\theta-2\pi (r_0^M+\frac{h}{2\pi}\theta)\sin\theta\big], \Delta_r = (r^M_0-r^m_0)
\end{split}
\label{straineq}
\end{equation}
where $\theta$, $r_0^M$ , $r_0^m$ and $h$ are the angle between the radial vector and the horizontal axis, major axis at $\theta = 0$ , minor axis at $\theta=0$ and interlayer spacing. The curvature is constantly changing over the NS surface with maxima (minima) at $\theta = \frac{n\pi}{2}$ for even (odd) $n$ (see supplementary section VI for details of the derivation of Eq. \ref{straineq} ~\footnotemark[\value{footnote}])\cite{landau1986theory,liu2014elastic}. This makes the strain landscape inhomogeneous for NS, which may lead to a similar kind of layer decoupling in the NS.

\begin{figure}[t]
    \centering
    \includegraphics[width = 0.9\linewidth]{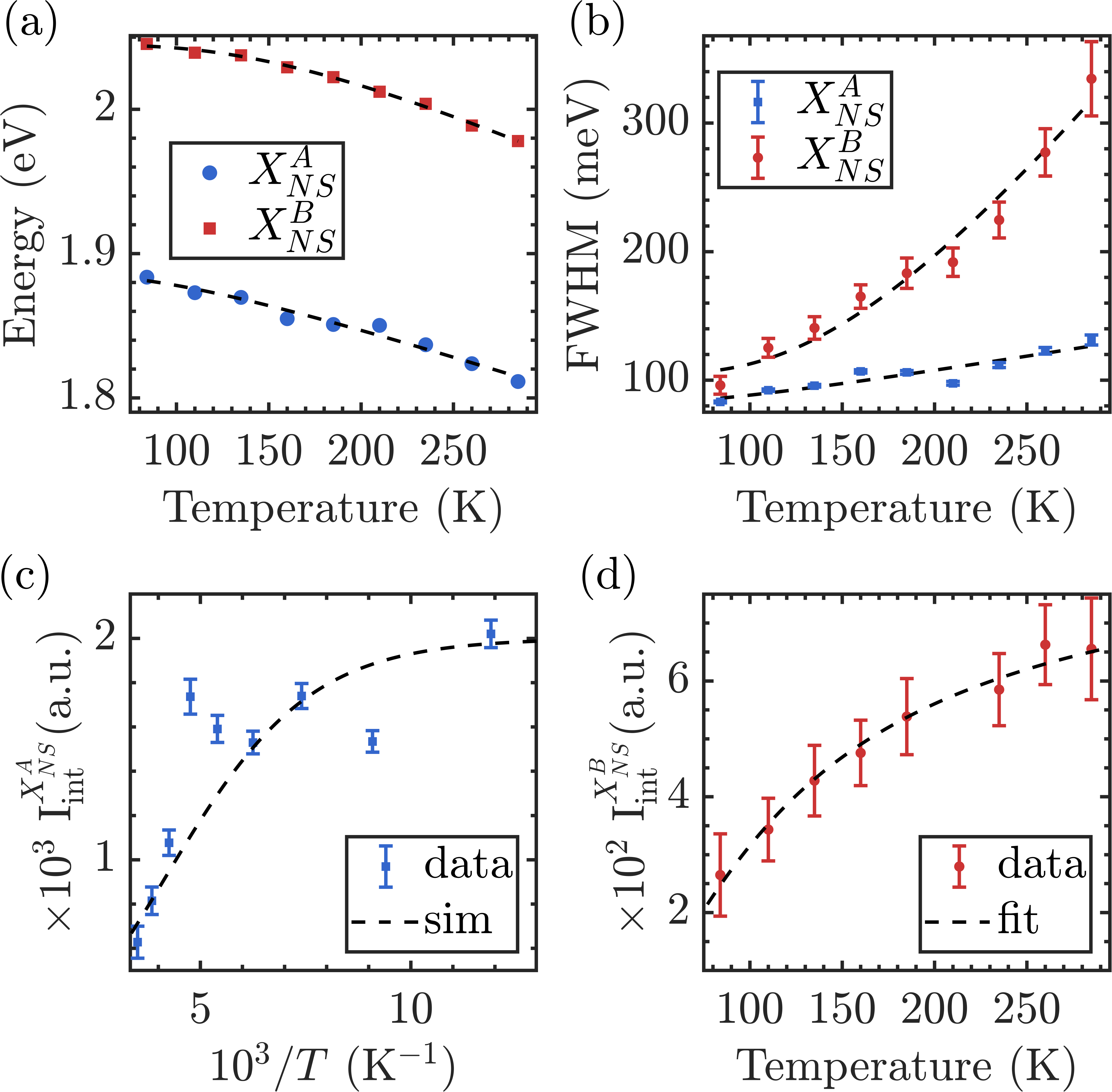}
   \caption{Phenomenological fitting of temperature-dependent excitonic peak positions (a) and temperature-dependent FWHM (b) for $\ans$ and $\bns$ (c) Integrated PL peak intensity for $\ans$ as a function of inverse temperature (d) Integrated PL peak intensity for $\bns$ as a function of temperature}
    \label{fig:4}
\end{figure}

An increase in interlayer separation can lead to lowering of EXPC for out-of-plane phonon mode \cite{PhysRevB.98.035411}.To extract the effect on EXPC of NS, we fitted the temperature-induced shift in the excitonic peak positions for $\ans$ and $\bns$ [FIG. \ref{fig:4}(a)] with the phenomenological O'Donnell and Chen's equation  \cite{o1991temperature}

\begin{equation}
    E(T) = E(0) - S\langle\hbar\omega\rangle\bigg(\coth\frac{\langle\hbar\omega\rangle}{2k_BT}-1\bigg)
    \label{eq:2}
\end{equation}
Here, $E(0)$ is the peak position at 0 K, $S$ is the Huang-Rhys factor which measures the strength of EXPC, and $\langle\hbar\omega\rangle$ is the average phonon energy\cite{helmrich2018exciton}. $S_\text{NS}$ for $\a$ is found to be 2.46, which is found to be closer to the previously reported value of monolayer \ch{MoS2}\cite{PhysRevB.93.165412}. $S$ is expected to be higher in NS due to lower dimensionality and the presence of multiple layers\cite{zhang2015valence,baldini2019exciton,PhysRevLett.114.136403}. The lowering of $S$ for $\ans$ signifies lower EXPC due to layer decoupling. For $\bns$ the value of $S$ is 3.75. For $\b$, the parameter is less affected as it couples less with the out-of-plane optical phonon mode $A_{1g}$\cite{saigal2015phonon,PhysRevLett.114.136403}. The value of $\langle\hbar\omega\rangle$ for $\ans$  $\sim$  24 meV, which is close to the previously reported value for exfoliated \ch{MoS2} ML \cite{saigal2015phonon,cadiz2017excitonic}.
For $\bns$ the value of $\langle\hbar\omega\rangle$ is $\sim$ 43 meV, which is quite close to the energy of the available phonon modes in \ch{MoS2} \cite{lee2010anomalous}. 

To estimate the EXPC-induced broadening of the excitonic peaks we used the expression derived in \cite{rudin1990temperature,srinivas1992intrinsic}, 
\begin{equation}
    \Gamma(T) = \Gamma_0 + \Gamma_{1}T + \frac{\Gamma_{2}}{e^{\frac{E_{ph}}{k_BT}}-1},
    \label{eq:3}
\end{equation}
to fit temperature-dependent full width at half maxima (FWHM) for different excitonic peaks [FIG.  \ref{fig:4}(b)]. In this equation, $\Gamma_0$ is the intrinsic linewidth, $\Gamma_1$ is the linewidth due to exciton-acoustic phonon coupling, and the last term arises due to exciton-optical phonon coupling. The value of $E_{ph}$ is taken equal to the value of $\langle\hbar\omega\rangle$ extracted by fitting Eq. \ref{eq:2}. In our study, the second term is very small ($\sim40-70 \   \mu eV/K$) for all the concerned excitonic peaks, which matches well with previous studies \cite{dey2016optical,moody2015intrinsic,cadiz2017excitonic}. $\Gamma_2$ for $\ans$ ($\sim 49\ \text{meV}$), which is close to the earlier reported value for exfoliated \ch{MoS2} ML \cite{cadiz2017excitonic}.
This is in good agreement with the lowered EXPC strength in the NS system. All parameters from fitting Eq. \ref{eq:2} and Eq. \ref{eq:3} are listed in supplementary material section VIII in TABLE S1 and TABLE S2 respectively ~\footnotemark[\value{footnote}].

As mentioned in the previous section, the PL from NS can be immune to layering-induced bandgap reduction and the consequential red-shift \cite{huang2020transition}. In fact, at low temperatures, we can spot an overall blueshift in the peak positions of $\ans$ and $\bns$ with respect to $\aml$ and $\bml$ (FIG.  \ref{fig:1}(c)). From the fitting of Eq. \ref{eq:2}, $E(0)$ for both $\a$ and $\b$ is blue-shifted by approximately 20 meV and 10 meV, respectively, in NS compared to ML.  This is very unusual, as the multilayer structure and strain in the system should cause a redshift of the peak position \cite{splendiani2010emerging, zhang2015valence, conley2013bandgap, shi2013quasiparticle}. Previously, a similar blueshift was reported in folded 1H-\ch{MoS2} \cite{crowne2013blueshift}. 

Structurally, NS possesses few similarities with the folded 1H-\ch{MoS2}, such as broken inversion symmetry and increased interlayer separation, which can result in a similar blueshift in NS due to a similar modulation of Coulomb interactions. Anisotropic and weak screening occurs in the direction perpendicular to the plane in a 2D layer \cite{chernikov2014exciton}. As a result, the Coulomb fields are intensified in the out-of-plane direction as opposed to the in-plane direction. 

Due to the interlayer separation, the electrons and holes created in the exciton annihilation process cannot move from one layer to another. Especially at lower temperatures, phonon-mediated scattering processes that can help non-tunneling transfer of charges through the connected region of the scroll, drop significantly. This results in increased in-plane screening and further reduces the exciton binding energy\cite{crowne2013blueshift}, causing the blueshift of the exciton peak positions.

Next, we have studied the temperature-dependent change in the integrated intensity of $\ans$ and $\bns$ [FIG. \ref{fig:4}(c-d)].  The Arrhenius plot of $\ans$ [FIG.  \ref{fig:4} (c)] shows a decay in a seemingly activated manner in the high-temperature region with slope tending to an activation energy $E_A$ and saturation at lower temperatures. This behaviour is simulated using the following equation\cite{saigal2015phonon}  
\begin{equation}
    I = I_0/[1+\gamma/(e^{\frac{E_A}{k_BT}}-1)]
\end{equation}
with $E_A = 50 \,\, \rm{meV}$ is taken equal to the relevant phonon mode's ($A_{1g}$) energy \cite{saigal2016evidence} (see supplementary section IX for Raman spectroscopy details) ~\footnotemark[\value{footnote}]) and $\gamma = 12$. This model captures, through the constant $\gamma = \tau_R/\tau_{NR_{0}}$, the contribution of excited phonons, where $1/\tau_{NR_{0}}$ and $1/\tau_{R}$ are the non-radiative and radiative decay rate of the exciton respectively. The fairly low value of $\gamma$ \cite{saigal2015phonon, saigal2016evidence} for $\ans$ is indicative of the layer decoupling and effective lowering of coupling between $\ans$ and $A_{1g}$ phonons. 

 $\bns$ shows a steady decline in integrated PL peak intensity with lowering of temperature [FIG.  \ref{fig:4}(d)]. Whereas earlier reports claim that for $\b$ in ML \ch{MoS2}, the integrated intensity doesn't change much with temperature \cite{saigal2015phonon}. This trend is also observed in the case of the ML under study (see supplementary section X FIG. S12 ~\footnotemark[\value{footnote}]). But in a recent study on ML \ch{MoS2} and \ch{MoSe2}, the reduction in $\b$  intensity with decreasing temperature has been reported \cite{katznelson2022bright}. For ML \ch{MoS2}, the spin forbidden dark-$\b$ state lies $\sim$20 meV below the bright-$\b$ state as theoretically calculated in ref.\cite{echeverry2016splitting}. Experimentally this value is determined to be $\sim$14 meV \cite{robert2020measurement}. In semiconductors with such excitonic states, where spin-forbidden dark excitonic states lie below bright excitonic states, the temperature-dependent emission intensity from the bright state can be expressed as \cite{zhang2015experimental}:
\begin{equation}
    I = I_0/(1+e^{\Delta/k_BT})
    \label{eq.db}
\end{equation}
Here, $\Delta$ is dark-bright splitting energy, $k_B$ is the Boltzmann constant, and $I_0$ is the proportionality constant. Fitting Eq. \ref{eq.db} we have found out $\Delta \sim 13$ meV, which is close to previously determined value for the dark-bright splitting \cite{robert2020measurement}.  The lowering of phonon-assisted dark-to-bright conversion in lower temperatures can explain the diminished population of $\b$ states in \ch{MoS2} NS at lower temperatures. 

In summary, we have demonstrated that the layers in \ch{MoS2} nanoscrolls decouple due to misalignment and non-uniform curvature by investigating the interlayer interactions through PL spectroscopy. Our findings reveal that the decoupling of layers and the effective detachment from the substrate cause these NS to behave like a collection of suspended ML \ch{MoS2}.  This unique scrolled topology provides a novel platform for exploring the optoelectronic properties of \ch{MoS2} in a quasi-one-dimensional geometry. 

Furthermore, the misalignment of layers in NS gives rise to interesting moir\'e-like patterns,  which can be a promising subject for future investigations. These patterns, resulting from the relative twist between layers, may lead to the emergence of novel electronic and optical properties in NS. 

Our study establishes a valuable platform for gaining a deeper understanding of the modifications in the band structure and the associated trends in \ch{MoS2} NS, driven by the inherent structural complexity of the NS. Our results underscore the critical role of stacking order control and scroll topology in tailoring the optoelectronic properties of \ch{MoS2} NS. This is expected to open up new avenues for research and potential applications in two-dimensional materials.


\textit{Acknowledgments.}  A.M. acknowledges the support of the European Research Council (ERC) under the European Union's Horizon 2020 research and innovation program (Grant Agreement No. 865590) and the Research Council UK [BB/X003736/1]. A.R. acknowledges funding support from DST SERB Grant no. CRG/2021/005659 and partial funding support under the CEFIPRA, project no. 6104-2. This research used electron microscopy facilities of the Center for Functional Nanomaterials (CFN), which is a U.S. Department of Energy Office of Science User Facility at Brookhaven National Laboratory under Contract No. DE-SC0012704. The authors would like to acknowledge funding from  NM-ICPS of the DST, Govt. Of India through the I-HUB Quantum Technology Foundation, Pune, India. S. C. and T. C. thank the PMRF, Govt. of India for providing research fellowship. T. C. also thanks the Commonwealth Scholarship Commission and the FCDO in the UK for the Split-site fellowship.

S. C. and T. C. have contributed equally to this work.

\bibliographystyle{apsrev4-2}
\bibliography{bib}

\par
\end{document}


\title{Supplementary Material: Effects of Reduced Interlayer Interactions on the K-point Excitons of \ch{MoS2} Nanoscrolls}

\author{Sagnik \surname{Chatterjee}}
\email{chatterjee.sagnik@students.iiserpune.ac.in}
\thanks{The author contributed equally to this work.}
\affiliation{Department of Physics, Indian Institute of Science Education and Research (IISER), Pune 411008, India}
\author{Tamaghna \surname{Chowdhury}}
\email{tamaghna.chowdhury@students.iiserpune.ac.in}
\thanks{The author contributed equally to this work.}
\affiliation{Department of Physics, Indian Institute of Science Education and Research (IISER), Pune 411008, India}
\affiliation{Department of Physics and Astronomy, University of Manchester, United Kingdom M13 9PL}
\author{Pablo D\'iaz \surname{N\'u\~nez}}
\affiliation{Department of Physics and Astronomy, University of Manchester, United Kingdom M13 9PL}
\affiliation{National Graphene Institute, University of Manchester, United Kingdom, M13 9PL.}
\author{Nicholas \surname{Kay}}
\affiliation{Department of Physics and Astronomy, University of Manchester, United Kingdom M13 9PL}
\affiliation{National Graphene Institute, University of Manchester, United Kingdom, M13 9PL.}
\author{Manisha \surname{Rajput}}
\affiliation{Department of Physics, Indian Institute of Science Education and Research (IISER), Pune 411008, India}
\author{Sooyeon \surname{Hwang}}
\affiliation{Center for Functional Nanomaterials, Brookhaven National Laboratory, Upton, NY 11973, USA }
\author{Ivan \surname{Timokhin}}
\affiliation{Department of Physics and Astronomy, University of Manchester, United Kingdom M13 9PL}
\affiliation{National Graphene Institute, University of Manchester, United Kingdom, M13 9PL.}
\author{Artem \surname{Mishchenko}}
\email{artem.mishchenko@gmail.com}
\affiliation{Department of Physics and Astronomy, University of Manchester, United Kingdom M13 9PL}
\affiliation{National Graphene Institute, University of Manchester, United Kingdom, M13 9PL.}
\author{Atikur \surname{Rahman}}
\email{atikur@iiserpune.ac.in}
\affiliation{Department of Physics, Indian Institute of Science Education and Research (IISER), Pune 411008, India}
\date{\today}

\maketitle

\def\ans{X_{NS}^A}
\def\bns{X_{NS}^B}
\def\aml{X_{ML}^A}
\def\bml{X_{ML}^B}
\def\ians{I_{X_{NS}^A}}
\def\ibns{I_{X_{NS}^B}}
\def\iaml{I_{X_{ML}^A}}
\def\ibml{I_{X_{ML}^B}}
\def\a{X^A}
\def\b{X^B}
\def\ia{I_{X^A}}
\def\ib{I_{X^B}}
\def\ms{\mathrm{MoS}_2}

\section{Experimental details}
\subsection{$\ms$ monolayer synthesis}
$\ms$ ML are synthesized using standard chemical vapour deposition (CVD) technique using a single zone furnace. MoO$_3$ and Sulfur powders in a 1:80 ratio, kept at separate boats at 18 cm apart, are used as precursors for the growth. The growth occurs in 300 nm SiO$_2$ coated Si substrate kept at a distance of 5 cm from the MoO$_3$ boat. A steady flow of Ar at 20 sccm is maintained during the entire period of the process to provide an inert atmosphere for growth.  The temperature during the reaction is set at 973 K and the reaction occurs for 10 min. 
\subsection{Nanoscroll preparation}
$\ms$ NSs are formed by dropping ethanol aqueous solution (2:1) on freshly prepared $\ms$ ML and letting it evaporate naturally as described in ref \cite{cui2018rolling}. CVD is a high-temperature synthesis process and the reaction to synthesize $\ms$ happens at 973 K. After the reaction when the substrate containing the $\ms$ ML cooled down then due to the difference between thermal expansion coefficient of $\ms$ ML and the SiO$_2$ coated Si substrate, strain builds up in the system. When ethanol aqueous solution is dropped onto the substrate, the strain is released during evaporation of the solution. The solution penetrates between the substrate and the $\ms$ ML eventually rolling up the ML.
\subsection{Temperature Dependent Photoluminescence Spectroscopy}
The temperature-dependent PL measurements were done using continuous flow Oxford Instruments optical cryostat HiRes. The PL spectra were obtained in a WiTec Alpha 300 spectrometer mounted with a grating of 600 lines per mm using a Nikon 100$\times$objective (numerical aperture (N. A.) $= 0.80$) lens. The samples were excited with a continuous wave laser of wavelength($\lambda$) 514.5 nm. A water-cooled charged couple device (CCD) was used for detection.  All measurements were done at a low pressure of $2\times10^{-5}$ mbar pressure. The spectral resolution of PL spectra is 0.5 meV. Liquid nitrogen was used as cryogen and temperature was varied between 84 K and 300 K. All measurements were done at an excitation power less than equal to 1 mW to prevent laser heating-related damage to the sample. PL mappings were done at room temperature and ambient pressure by placing the sample on a piezostage using a Nikon objective lens of 100$\times$ magnification and 0.95 N. A. The laser spot size(d) was calculated using the formula $d = \frac{1.22\lambda}{N. A.}$\cite{francis1950fundamentals}

\section{Room temperature PL peak position mapping}
 \begin{figure}[H]
    \centering
    \includegraphics[width=1\linewidth]{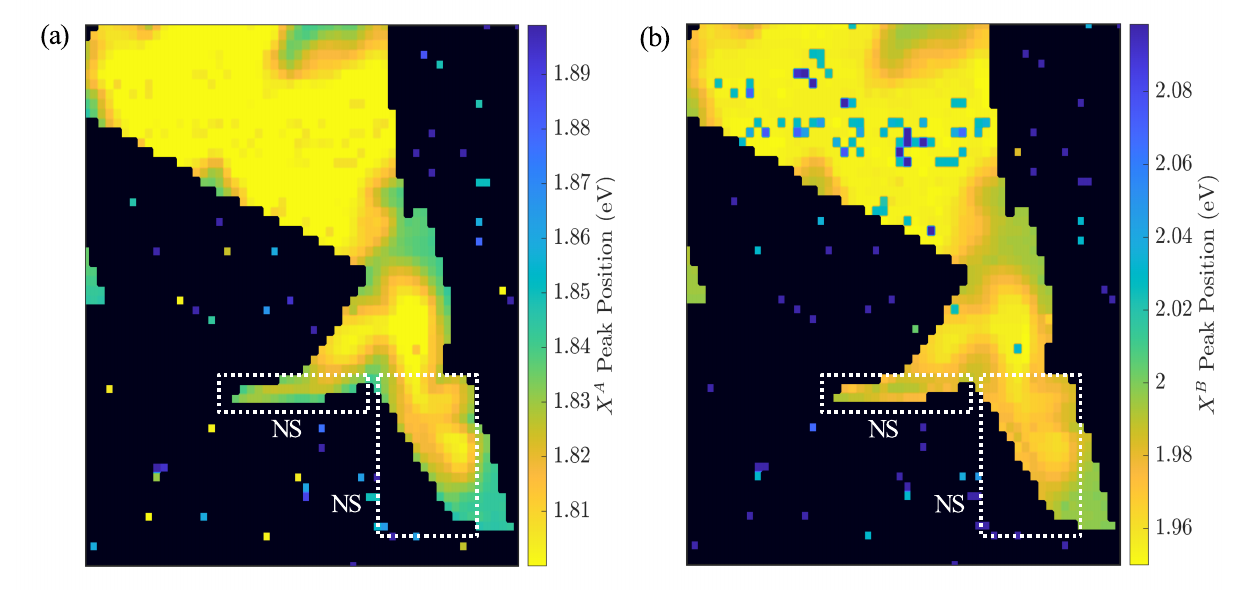}
    \caption{PL mapping showing peak position of $\a$ (a) and $\b$ (b). NS region is shown inside the square box with dashed lines.}  
    \label{fig:S8}
\end{figure}

 \section{Layer Stacking in 2H TMDs}
 In AA stacking similar atoms are on top of each other (M on M and X on X) and in AB stacking dissimilar atoms are on top of each other (M on X and X on M). The other stackings are of Bernal kind denoted by B in our notation. B$^{\rm{X/M}}$ denotes a stacking where the X atom of the top layer is directly above the M atom from the bottom layer and so on. B$^{\rm{X/M}}$ and B$^{\rm{M/X}}$ are equivalent stackings, yielding five non-equivalent stacking for 2H TMDs. \cite{naik2019kolmogorov}
 \begin{figure}[H]
    \centering
    \includegraphics[width=\linewidth]{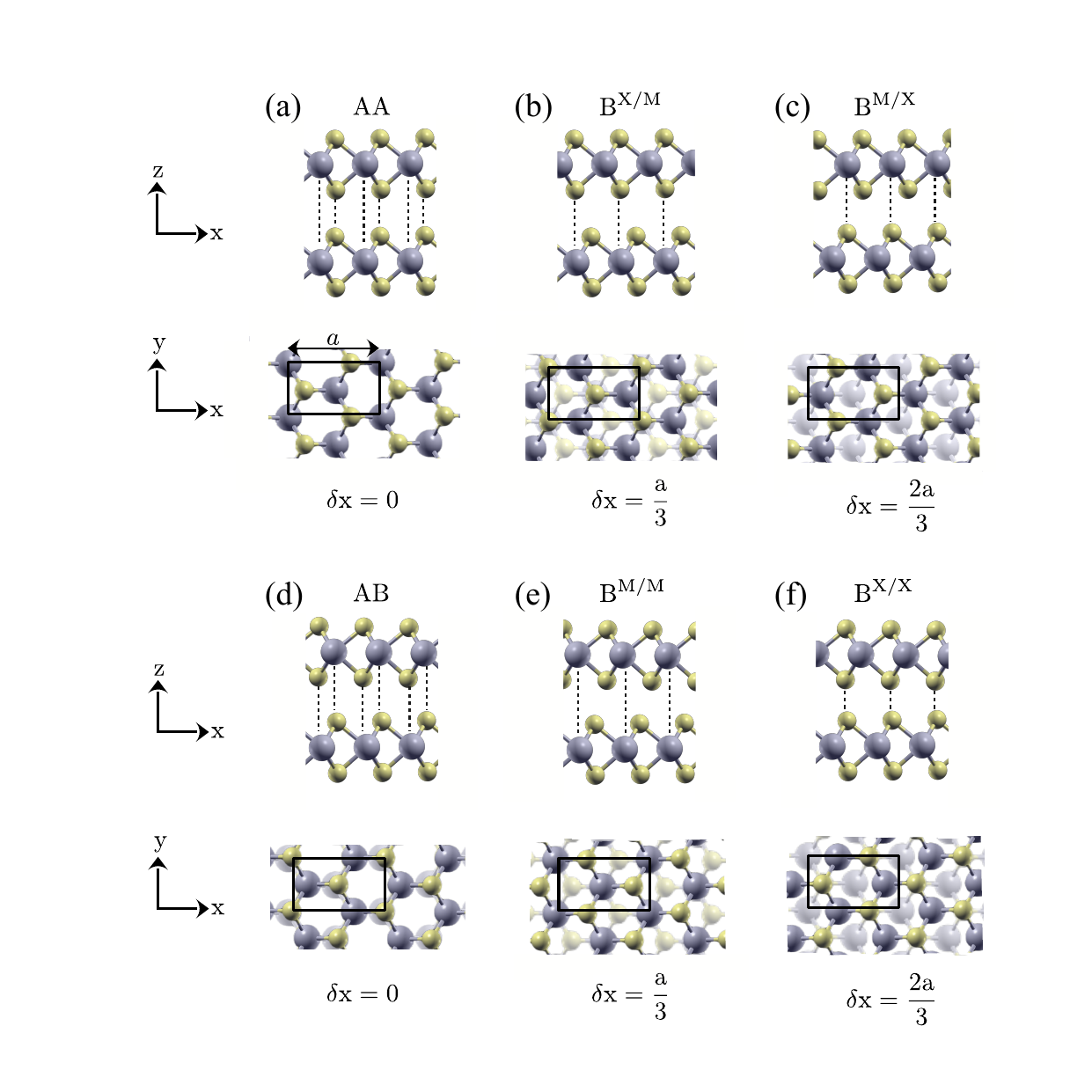}
    \caption{High symmetry stacking of bilayer 2H TMD obtained by sliding the top layer with respect to the bottom layer. The orthogonal unit cell is marked. Starting with the AA stacking ($\delta x = \delta y = 0$) (a), sliding by $\delta x = a/3$  leads to the B$^{\rm{X/M}}$ stacking (b) and sliding by $\delta x = 2a/3$  leads to the B$^{\rm{M/X}}$ (c). Similarly, starting from AB stacking (d) yields B$^{\rm{M/M}}$ stacking (e) and B$^{\rm{X/X}}$ stacking (f). The yellow and blue spheres indicate S and Mo atoms, respectively.}
    \label{fig:S12}
\end{figure}

\section{HRTEM \& HAADF-STEM}

\subsection{HRTEM}
For HRTEM imaging NS from \ch{SiO2} coated \ch{Si} substrate are transferred to an amorphous carbon-coated Cu grid using a polystyrene based wet transfer method. 20\% v/v polystyrene solution in toluene is used for the process. The HRTEM imaging is performed using JEOL JEM-2200FS with an accelerating voltage of 200 kV.
\begin{figure}[H]
    \centering
    \includegraphics[width=\linewidth]{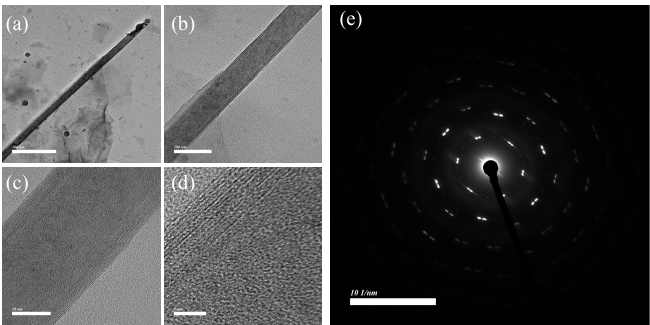}
    \caption{(a)HRTEM image of NS. Zoomed-in image in various magnifications is given in (b-d). scale bars for (a),(b),(c) and (d) are 500 nm, 100 nm, 20 nm and 5 nm respectively. (e) SAED taken on top of the NS showing mutually twisted layers with maximum angular spread of $\sim$ 30$^{\circ}$}
    \label{fig:S2}
\end{figure}

\subsection{HAADF-STEM}
 High-angle annular dark-field scanning transmission electron microscopy (HAADF-STEM) images were acquired with Hitachi HD2700C dedicated STEM with Cs probe corrector with an accelerating voltage of 200 kV.

The interlayer spacing is calculated along the line taking N  number of peaks as, \par

\vspace{0.1cm}   $ \frac{\rm{First \ Peak} -\rm{Last \  Peak} }{\rm{N} -1} = \frac{5.6-0.517}{8} \ \rm{nm} = 0.635 \ \rm
{nm} $
\begin{figure}[H]
    \centering
    \includegraphics[width=0.9\linewidth]{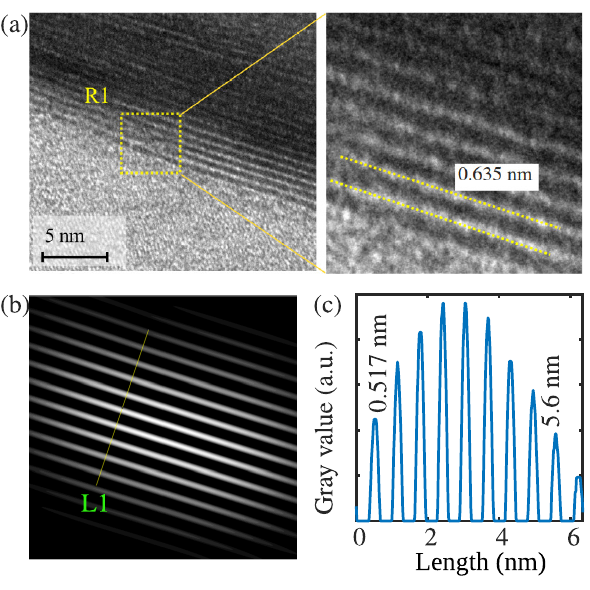}
    \caption{(a) TEM image of NS. R1 is zoomed in showing interlayer spacing of 0.635 nm (b) Cleared TEM image using FFT-Inverse FFT method (c) Gray value plot along line L1 in (b)}
    \label{fig:S2}
\end{figure}

\section{Atomic Force Microscopy}
\begin{figure}[H]
    \centering
    \includegraphics[width=1\linewidth]{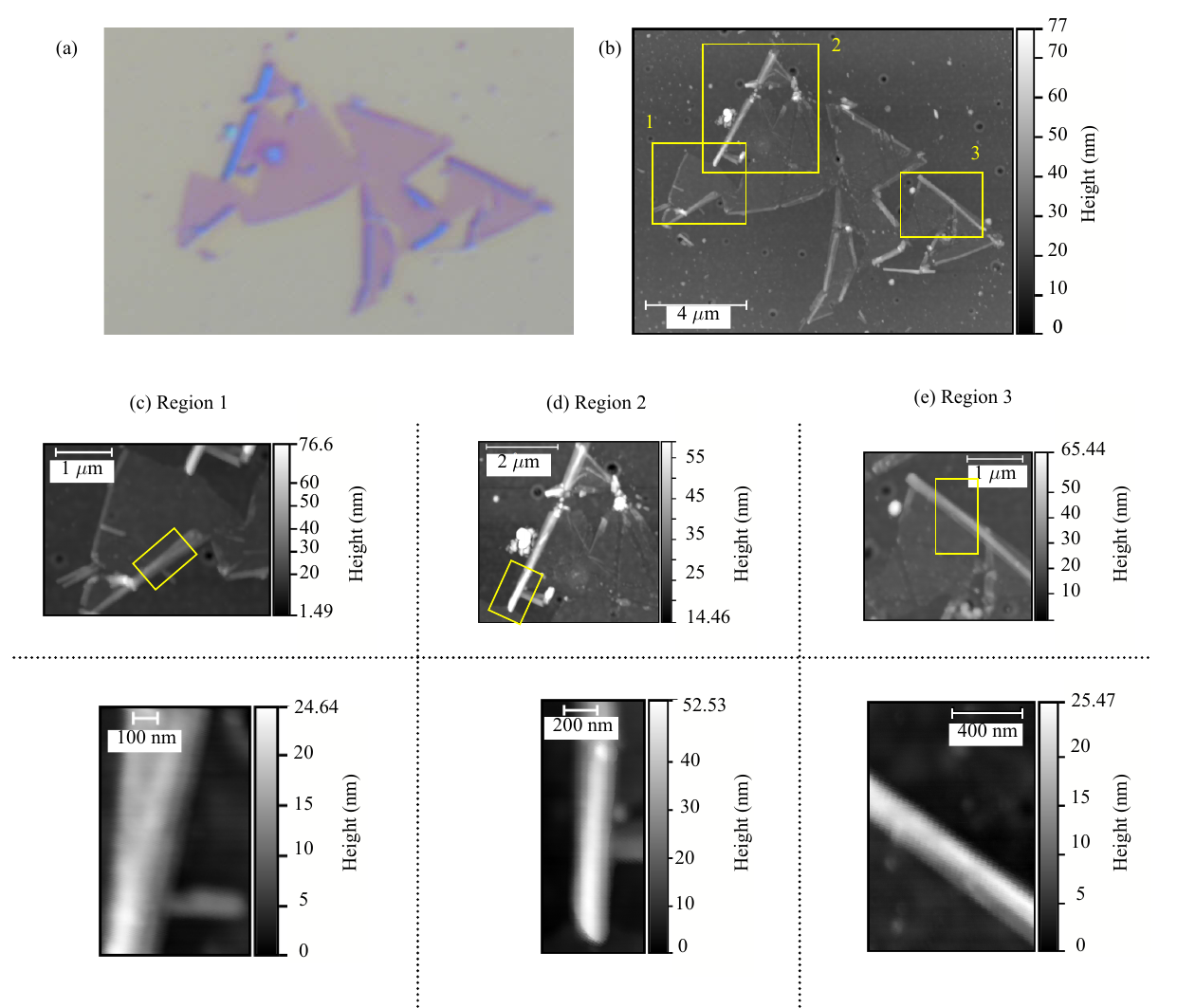}
    \caption{(a) Optical image of partially scrolled $\ms$ ML (b) AFM image of the partially scrolled $\ms$ ML shown in (a). (c-e) are AFM images of the regions marked as 1-3 in (b). The lower panel of (c-e) shows the AFM image for the marked region in the upper panel (c-e) }
    \label{fig:S3}
\end{figure}

\vspace{-0.3cm}
AFM images were 
acquired in a tapping mode using a commercial AFM microscope from Neaspec, GmbH. We used Pt$/$Ir coated AFM tips (Arrow$^{\rm{TM}}$ NCPt, nanoWorld) oscillating at its resonance frequency ($\Omega\sim 285$ kHz) with a tapping amplitude of about 60 nm.
AFM was employed to confirm the elliptical cross-section of the NS. The height of the NS is much smaller compared to the width as can be seen in FIG. \ref{fig:S3} (c-e) bottom panel. The overall structure of these NS resembles an Archimedean spiral shape with positive eccentricity.

\section{Structure of NS \& Strain} 
The ideal NS structure can be modelled as an Archimedean spiral defined by 
\begin{equation}
    r = r_0 + \frac{h}{2\pi}\theta
    \label{S2}
\end{equation}
where $r_0$ is the inner radius of the NS, $h$ is the interlayer separation and $\theta$ is the rotation angle with respect to the negative y-axis. The total length of the NS with $n$ number of layers will be 
\begin{equation}
  l =  2\pi r_0n + h\pi n^2
    \label{S3}
\end{equation}
In this description, the cross-section of the NS will be a spiral with eccentricity $e = 0$. In the case of real NS, the cross-section has either $e>0$ (oblate cross-section) or $e<0$ (prolate cross-section). In an experimental NS system the cross-section is likely to be oblate as confirmed by FESEM and AFM measurement and also reported in previous studies \cite{cui2018rolling,qian2020chirality}. This shape can be described as 
\begin{equation}
    r_{\rm{M}} = r_0^{\rm{M}} + \frac{h}{2\pi}\theta
      \label{S4}
\end{equation}
\begin{equation}
    r_{\rm{m}} = r_0^{\rm{m}} + \frac{h}{2\pi}\theta
      \label{S5}
\end{equation}
\begin{equation}
    x = r_{\rm{M}} \cos\theta
      \label{S6}
\end{equation}
\begin{equation}
    y = r_{\rm{m}} \sin\theta
      \label{S7}
\end{equation}
where $r_0^{\rm{M}}$ and $r_0^{\rm{m}}$ are the inner major and minor axis length respectively. $(x,y)$ is the coordinate of a point on the arc. The eccentricity is defined by $e = \frac{r_0^{\rm{M}}-r_0^{\rm{m}}}{r_0^{\rm{M}}+r_0^{\rm{m}}}$. \par
\begin{figure}[H]
    \centering
    \includegraphics[width=\linewidth]{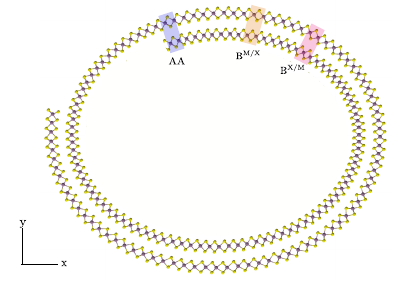}
    \caption{Cross-section of a NS with $e=0.2$, $h=0.65$ nm generated with Eq. \ref{S4}-\ref{S7}. The yellow and blue spheres indicate S and Mo atoms, respectively. Different possible stacking of neighbouring layers is highlighted in the figure.}
    \label{fig:S13}
\end{figure}

The two nonzero component of strain tensor, $u_{xx}$ and $u_{zz}$ for the cross-section can be defined as \cite{landau1986theory},
\begin{equation}
    u_{xx} = -t\frac{\partial^2 y}{\partial x^2}
      \label{S8}
\end{equation}
\begin{equation}
    u_{zz} = \frac{\eta t}{1 - \eta}\frac{\partial^2 y}{\partial x^2}
    \label{S9}
\end{equation}
    where, $t$ is the thickness of the ML and $\eta$ is the Poisson's ratio of the ML. For $\ms$ ML $t = 0.65\ \rm{nm}$ and $\eta = 0.25$\cite{liu2014elastic}. For NS using Eq. \ref{S2}-\ref{S7} 
\begin{equation}
    \frac{\partial^2 y}{\partial x^2} = 4\pi \frac{h^2+2\pi^2 r_mr_M - \frac{1}{2}\pi h \sin2\theta(r^M_0-r^m_0)}{(h\cos\theta-2\pi r_M\sin\theta)^3}
    \label{S10}
\end{equation}

From Eq. \ref{S10} it can be realized that the curvature on the surface of the NS is continuously varying with $\theta$ and thus the strain is inhomogeneous. The absolute value of curvature for $\theta = 0$ 
\begin{equation}
    \bigg|4\pi \frac{h^2+2\pi^2 (r_0^m+\frac{h}{2})(r_0^M+\frac{h}{2})}{h^3}\bigg|
\label{cur0}
\end{equation}
and for $\theta=\pi/2$ will be,
\begin{equation}
    \bigg|4\pi \frac{h^2+2\pi^2 (r_0^m+\frac{h}{4})(r_0^M+\frac{h}{4})}{8\pi^3(r_0^M+\frac{h}{4})^3}\bigg|
    \label{curpiby2}
\end{equation}
Numerator of Eq. \ref{cur0} $>$ Numerator of Eq. \ref{curpiby2} and   Denominator of eq. \ref{cur0} $<$ Denominator of eq. \ref{curpiby2} as $r_0^M \gg h$. So curvature is highest at $\theta = 0$ and it decreases with increasing $\theta$ with minima at $\theta = \pi/2$. This same analysis can be extended to $\theta = \frac{n\pi}{2}$ where curvature will be maximum (minimum) for n even (odd).

This scenario is thus similar to cylindrically curved bilayer $\ms$ where the layers effectively decouple due to this inhomogeneity in strain\cite{koskinen2014density}.

\section{Fitted PL spectra for different temperatures}

\begin{figure}[H]
    \centering
    \includegraphics[width=\linewidth]{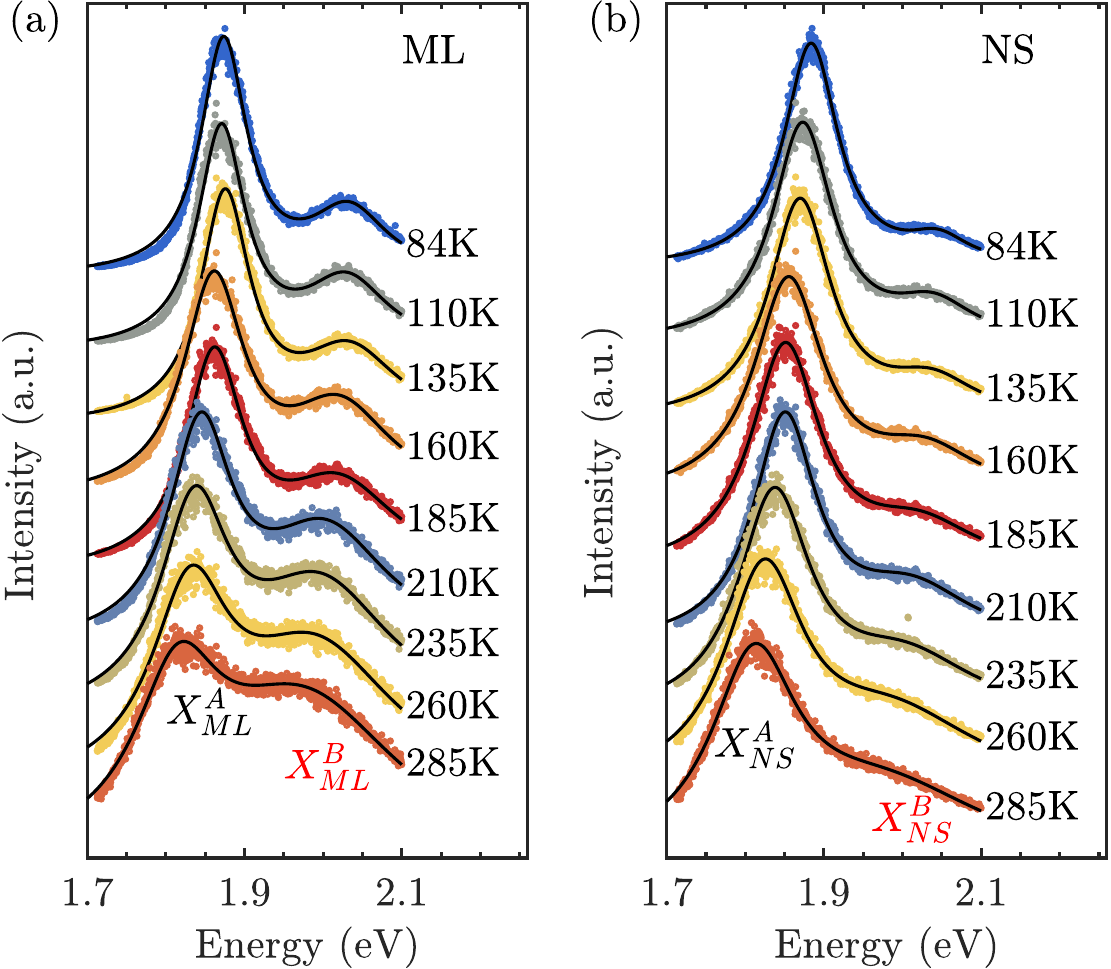}
    \caption{Fitted PL lineshape with two Lorentzian Function for $\ms$ (a)ML and (b)NS. The colored points are experimental data points and the black line is the fitted lineshape.}
    \label{fig:S4}
\end{figure}
\section{Semi-emperical fitting of peak position and FWHM for $\aml$ and $\bml$}
\begin{figure}[H]
    \centering
    \includegraphics[width=0.8\linewidth]{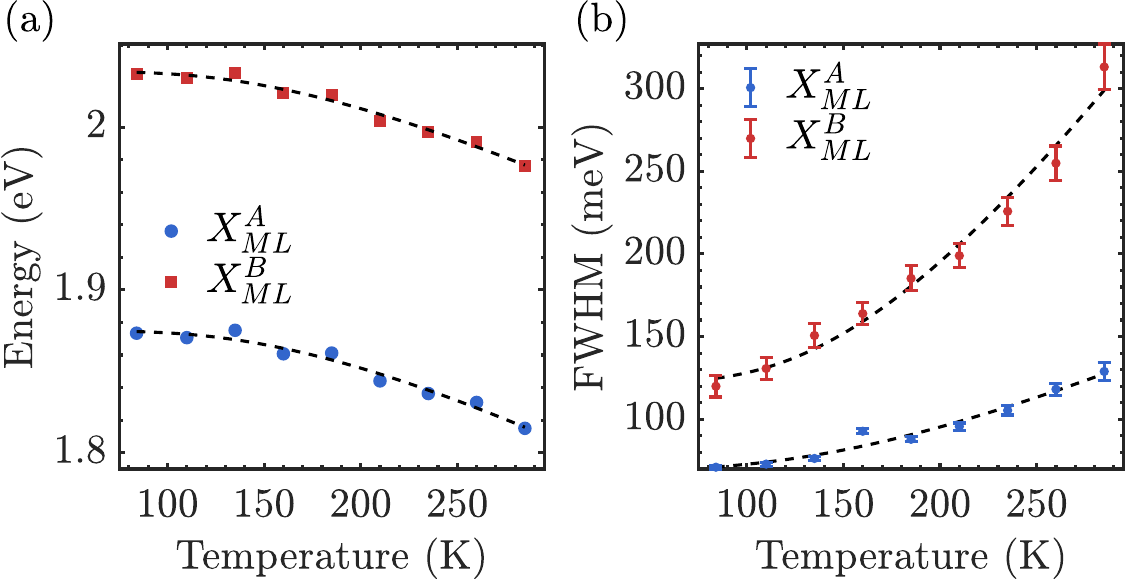}
    \caption{Phenomenological fitting of temperature-dependent excitonic peak positions (a) and FWHM (b) for $\aml$ and $\bml$}
    \label{fig:S5}
\end{figure}
{Value of parameters from temperature-dependent PL peak position and FWHM fit}
\begin{table}[H]
    \centering
    \begin{tabular}{|c|c|c|c|}
    \hline
               & $E(0) (eV)$  & $S$ & $\langle\hbar\omega\rangle$ (meV) \\
               \hline
        $\ans$ &1.89 & 2.46 & 24.2   \\
        $\aml$ &1.87 & 3.96 & 50.3    \\
        $\bns$ &2.04& 3.67 &  43.7  \\
        $\bml$ &2.03& 3.75 &  48.5  \\
        \hline
    \end{tabular}
    \caption{Fitting parameters for temperature-dependent peak position of excitons fitted with Eq. 2 in main text}
    \label{tab:1}
\end{table}

\begin{table}[H]
    \centering
    \begin{tabular}{|c|c|c|c|}
    \hline
            &$\Gamma_I$ (meV) & $\Gamma_{LA}$ ($\mu$ev/K)&  $\Gamma_{LO}$ (meV) \\
            \hline
       $\ans$  &78.72  & 64 &49.55 \\
       $\aml$ &65.12  & 68 & 290.37\\
       $\bns$  &101.89  &44  &1001.2 \\
       $\bml$ &118.27  & 62 &1011.5 \\
       \hline
    \end{tabular}
    \caption{Fitting parameters for temperature-dependent FWHM of excitonic peaks fitted with Eq. 3 in main text}
    \label{tab:2}
\end{table}

\section{Raman spectroscopy of NS}
 The low-temperature Raman measurement was done using continuous flow Oxford Instruments optical cryostat
HiRes. The Raman spectra were obtained in a WiTec Alpha 300 spectrometer mounted with a grating of 1800 lines per mm
using a Nikon 100X objective (numerical aperture (N. A.) = 0.80) lens. The samples were excited with a continuous wave laser of wavelength 514.5 nm and the spot diameter was $\sim$ 0.4 $\mu$m.
 Raman mappings were performed at room temperature with a 532 nm laser excitation on a Horiba XploRA setup with a grating of 1800 lines per mm and a 100X objective. 
  \begin{figure}[H]
    \centering
    \includegraphics[width=0.8\linewidth]{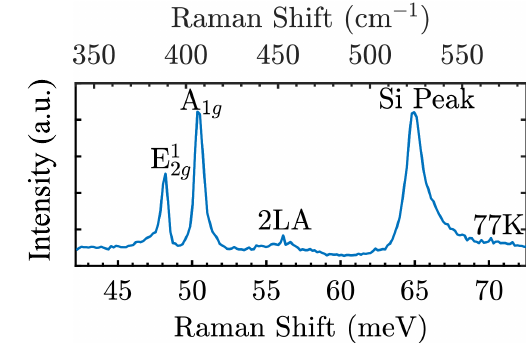}
    \caption{Raman Spectra of $\ms$ NS at 77K showing in-plane vibrational mode $\rm{E}^1_{2g}$ (388.8 cm$^{-1}$/48 meV) and out of plane vibrational mode $\rm{A}_{1g}$ (406 $\rm{cm}^{-1}$/50 meV) }  
    \label{fig:S9}
\end{figure}
 
  \begin{figure}[H]
    \centering
    \includegraphics[width=\linewidth]{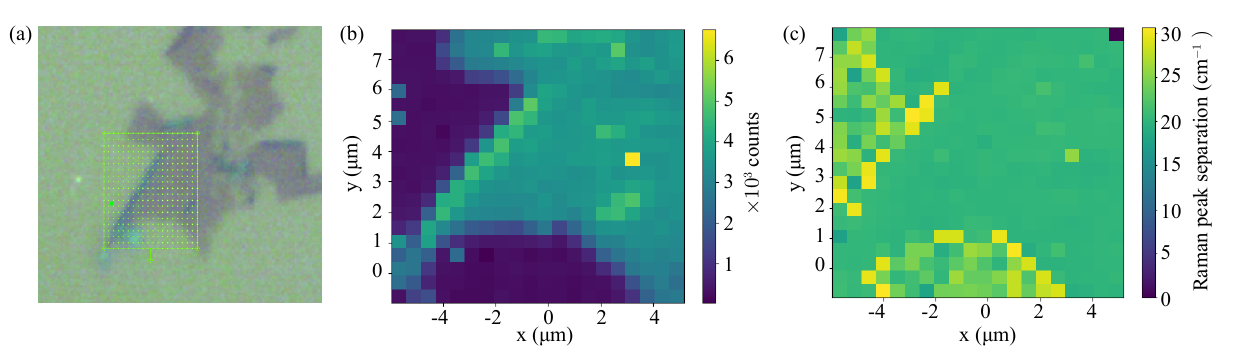}
    \caption{(a) Optical image of partially folded $\ms$ ML. (b) Room temperature mapping of the total count of  $\rm{E}^1_{2g}$ and $\rm{A}_{1g}$ peak region for the highlighted region shown in (a).  (c) Room temperature mapping of the distance between $\rm{E}^1_{2g}$ and $\rm{A}_{1g}$ peak position for the highlighted region shown in (a). The spectra for each pixel are fitted with two Lorentzian functions and a constant background.}  
    \label{fig:S10}
\end{figure}
 \begin{figure}[H]
    \centering
    \includegraphics[width=\linewidth]{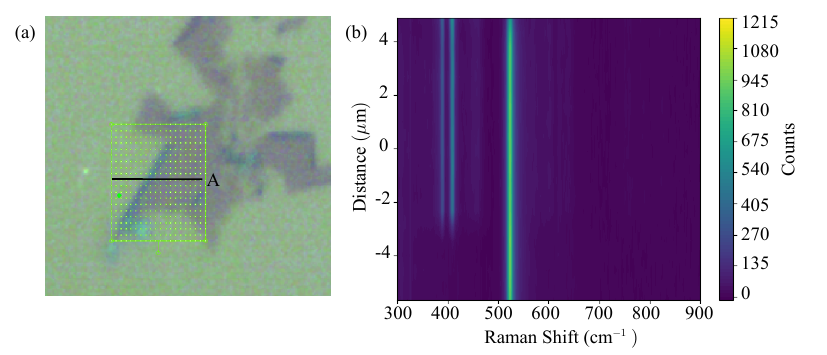}
    \caption{(a) Optical image of partially folded $\ms$ ML. (b) Room temperature scan of Raman spectra along line A in (a) Incident Polarization was circular and no polarization filter was used on detection.}  
    \label{fig:S11}
\end{figure}

\section{Temperature dependence of integrated PL peak intensity for $\aml$ and $\bml$}
\begin{figure}[H]
    \centering
    \includegraphics[width=0.5\linewidth]{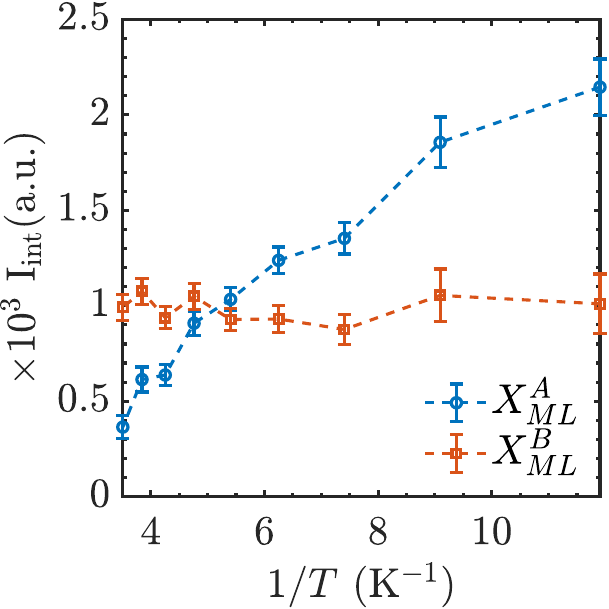}
    \caption{Integrated PL peak intensity for $\aml$ and $\bml$ as a function of inverse temperature}
    \label{fig:S6}
\end{figure}

\bibliographystyle{apsrev4-2}
\bibliography{bib.bib}